\documentclass[11pt]{article}
\usepackage{amssymb}

\parskip \baselineskip
\newcommand{\ds}{\displaystyle}
\newcommand{\R}{\mathbb{R}}

\newcommand{\Z}{\mathbb{Z}}

\newcommand{\ol}{\overline}

\newcommand{\ra}{\rightarrow}

\title{Computing the output distribution and selection probabilities of a stack filter from the DNF of its positive Boolean function}

\author{Marcel Wild\\[3pt]
 Department of Mathematical Sciences,
University of Stellenbosch\\
 Private Bag X1, Matieland 7602, South
Africa}
\date{}
\begin{document}
\maketitle

\begin{quote}
A{\scriptsize BSTRACT}: {\footnotesize Many nonlinear filters used in practise are {\it stack filters}.  An algorithm is presented which calculates the output distribution of an arbitrary stack filter $S$ from the disjunctive normal form (DNF) of its underlying positive Boolean function (PBF). Our algorithm avoids to enumerate the models of the PBF one by one, and thus is considerably more efficient than previous methods.  The so called rank selection probabilities can be computed along the way.}
\end{quote}

\section{Introduction}

Stack filters were invented in 1986 and have been a key topic of research in nonlinear signal processing ever since. Simply put, all aspects of a stack filter are reflected in its underlying positive Boolean function, and a basic familiarity of the latter concept is all that is required to understand this article. Using Google Scholar one can easily track the literature on various other aspects of stack filters, e.g. their output distribution.
In this article we present a new algorithm to calculate the output distribution. The new method, called stack filter $n$-algorithm, is an extension of the noncover $n$-algorithm [13] which generates, in compact form, all noncovers $X$ of given sets $A^\ast_1, \ldots, A^\ast_h$ (i.e. $X \not\supseteq A^\ast_i$ for all $1 \leq i \leq h$).

The stack filter $n$-algorithm is introduced by means of a medium-size example in Section 2. Section 3 is dedicated to its theoretic assessment. Section 4 touches upon five related matters, among which a numeric evaluation, and extensions of the stack filter $n$-algorithm that deliver the telling selection probabilities of [8], respectively handle the balanced stack filters of [12].

\section{The stack filter $n$-algorithm}

Fix $m \geq 1$ and put $w := 2m+1$. Let $b : \{0,1\}^w \ra \{0,1\}$ be a positive Boolean function (PBF), i.e. one without negated variables. Refering to e.g. [2], an operator $S$ from $\R^\Z$ in itself defined by the $k$-th component of $Sz$ being
\begin{equation}
[S z]_k \quad : = \quad b(z_{k-m}, \ldots, z_k, \ldots, z_{k+m} ) \quad (k\in \Z)
\end{equation}

is called a {\it stack filter} of window size $w$ based on $b$. Notice that the PBF $b$ in (1) has been extended from $\{0,1\}^w \ra \{0,1\}$ to $\R^w \ra \R$ in the usual way, i.e. by replacing the logical connectives $\wedge$ and $\vee$ by the minimum respectively maximum operation for pairs of real numbers (while keeping the symbols). So, if
$$b(x_{-1}, x_0, x_1)\quad : = \quad ((x_0 \vee x_1)\wedge x_{-1}) \vee x_0 \qquad (x_i \in \{0,1\}),$$
then
$$b(3,2,4) = ((2\vee 4) \wedge 3) \vee 2= (4 \wedge 3) \vee 2 = 3 \vee 2 = 3.$$
By construction each stack filter $S$ is {\it translation invariant} in the sense that pushing the series $x$ ten units to the right and then applying $S$ yields the same as first applying $S$ and then pushing ten units to the right. So $S$ is completely determined by formula (1) for $k =0$.

Let $Z = (\ldots, Z_{-1}, Z_0, Z_1, \ldots )$ be a doubly infinite sequence of independent indentically distributed (i.i.d.) random variables. Let $F_Z(t)$ be their common (cumulative) distribution function, i.e. $F_Z (t) := Prob(Z_i \leq t)$ is the probability that $Z_i$ is at most $t$.  By translation invariance the {\it output distribution} $F_{SZ}(t) := Prob((SZ)_i \leq t)$ is independent of $i$. It is known that there is a well defined function $\phi_S(p)$, called the {\it distribution transfer} of $S$, such that
$$F_{SZ}(t)\quad =\quad \phi_S(F_Z(t)) \qquad (t \in \R).$$
What's more, $\phi_S(p)$ is a {\it polynomial} which can be calculated [15], [2, p.223 ] as
\begin{equation}
\phi_S(p) \quad = \quad \ds\sum_{b(x) =0} p^{|\mbox{Zero}(x)|} \cdot q^{|\mbox{One}(x)|}
\end{equation}
where $q:= 1-p$ and $b$ is as in (1). The summation is over all bitstrings $x \in \{0,1\}^w$ with $b(x) =0$, where by definition
$$\begin{array}{lll}
\mbox{Zero}(x) & := & \{1 
\leq i \leq w | \ x_i = 0 \}, \\
\\
\mbox{One} (x) &: = & \{ 1 \leq i \leq w | \ x_i =1 \}. \end{array}$$
For instance, consider this positive Boolean function $b_1$ which is already in disjunctive normal form (DNF). It is of type $\{0,1\}^9 \ra \{0,1\}$ but we like to scale as $\{0,1\}^W \ra \{0,1\}$ with $W: = \{-4,-3, -2, -1, 0, 1,2, 3, 4\}$:
\begin{eqnarray} b_1(x_{-4}, \ldots, x_4) & = &  (x_{-2} \wedge x_{-1} \wedge x_0) \vee (x_{-1} \wedge x_0 \wedge x_1) \vee (x_0 \wedge x_1 \wedge x_2) \\
\nonumber \\
 & \vee & (x_{-4} \wedge x_{-3} \wedge x_{-2} \wedge x_1 \wedge x_2 \wedge x_3 )
 \vee (x_{-3} \wedge x_{-2} \wedge x_{-1} \wedge x_1 \wedge x_2 \wedge x_3 ) \nonumber \\
\nonumber \\ 
& \vee & (x_{-3} \wedge x_{-2} \wedge x_{-1} \wedge x_2 \wedge x_3 \wedge  x_4 ). \nonumber 
\end{eqnarray}
In view of (2) we wish to encode the family $\mbox{Mod}$ of all $x =(x_{-4}, x_{-3}, \ldots, x_4)$ in $\{0,1\}^W$ with $b(x) = 0$ in a compact way\footnote{If we were to start with the {\it conjunctive} normal form (CNF) of $b$, we would end up with a compact representation of the set Mod$'$ of all $x \in \{0,1\}^W$ with $b(x) =1$. Hence, instead of (2), a dual kind of formula would yield $\phi_S(p)$.}. First note that
$$\mbox{Mod} \quad = \quad \mbox{Mod}_1 \cap \mbox{Mod}_2 \cap \mbox{Mod}_3 \cap \mbox{Mod}_4 \cap \mbox{Mod}_5 \cap \mbox{Mod}_6,$$
where the family $\mbox{Mod}_i$ corresponds to the $i$-th conjunction in (3). For instance we write
$$\mbox{Mod}_1 \quad := \quad \{x \in \{0,1\}^W | \ x_{-2} \wedge x_{-1} \wedge x_0 =0\}\quad = \quad (2,2,n,n,n,2,2,2,2)$$
because $x_{-2} \wedge x_{-1} \wedge x_0=0$ {\bf (nul)} if and only if {\it at least one of} $x_{-2}, x_{-1}, x_0$ {\it is nul}, and the other variables $x_{-4}, x_{-3}, x_1, x_2, x_3, x_4$
can independently assume the ${\bf 2}$ values $0$ and $1$. Thus $(1,1,0,1,0,1,0,1,1) \in \mbox{Mod}_1$  but $(0,0,1,1,1,0,0,1,0) \not\in \mbox{Mod}_1$. If we identify a $0,1$-string $x$ with the subset $X = \{ i \in W : x_i =1\}$ of $W$ then $\mbox{Mod}_1$ consists of all {\it noncovers} $X$ of $A^\ast_1: = \{-2, -1, 0\}$ in the sense that $X \not\supseteq A^\ast_1$. The noncover $n$-algorithm  from [13] (more on that in Section 3) generates all  simultaneous noncovers of the given sets (here deriving from the terms of a PBF) $A^\ast_1, A^\ast_2, \ldots, A^\ast_6$ as follows:

\hspace*{2cm} \begin{tabular}{|c|c|c|c|c|c|c|c|c|cl}
$-4$ & $-3$ & $-2$ & $-1$ & $0$ & $1$ & $2$ & $3$ & $4$ &  & \\ \hline
2 & 2& $n$ & $n$ & $n$ & 2 & 2& 2&2& \quad $PC =2$\\ \hline \end{tabular}

\hspace*{2cm} \begin{tabular}{|c|c|c|c|c|c|c|c|c|l} \hline
2 & 2& 2 & {\bf n} & {\bf n} & 2 & 2 & 2& 2& \quad $PC=3$\\ \hline
 2 & 2& 0 & {\bf 1} & {\bf 1} & 0 & 2 & 2 & 2& \quad $PC =3$\\ \hline \end{tabular}

\hspace*{2cm} \begin{tabular}{|c|c|c|c|c|c|c|c|c|l} \hline
2 & 2& 2 & 2 & {\bf 0} & 2 & 2 & 2 & 2& \quad $PC = 4$\\ \hline
2 & 2& 2& 0 & {\bf 1} & $n$ & $n$ & 2 & 2& \quad $PC = 4$\\ \hline
2 & 2 & 0 & 1 & 1& 0 & 2& 2 &2 & \quad $PC = 3$\\ \hline \end{tabular}

\hspace*{2cm} \begin{tabular}{|c|c|c|c|c|c|c|c|c|l} \hline
$n$ & $n$ & $n$ & 2 & 0 & $n$ & $n$ & $n$ & 2& \quad$PC = 5$\\ \hline
2 & 2 & 2 & 0 & 1 & $n$ & $n$ & 2 & 2& \quad $PC = 4$\\ \hline
2 & 2 & 0 & 1 & 1 & 0 & 2 & 2 & 2& \quad $PC = 3$\\ \hline \end{tabular} 

\hspace*{2cm} \begin{tabular}{|c|c|c|c|c|c|c|c|c|l} \hline
2 & {\bf n} & {\bf n} & 2 & 0 & {\bf n} & {\bf n} & {\bf n} & 2& \quad $PC = 6$\\ \hline
0 & {\bf 1} & {\bf 1} & 0 & 0 & {\bf 1} & {\bf 1} & {\bf 1} & 2  & \quad $PC = 6$\\ \hline
2 & 2& 2 & 0 & 1 & $n$ & $n$ & 2 & 2 & \quad $PC = 4$\\ \hline
2 & 2 & 0 & 1 & 1 &0 & 2 & 2 & 2 & \quad $PC = 3$\\ \hline \end{tabular}

\hspace*{2cm} \begin{tabular}{|c|c|c|c|c|c|c|c|c|l} \hline
2 & {\bf n} & {\bf n} & 2 & 0 & 2 & {\bf n} & {\bf n} & 2&\quad  \mbox{final}\\ \hline
2 & {\bf 1} & {\bf 1} & n & 0 & 0 & {\bf 1} & {\bf 1} & n& \quad \mbox{final}\\ \hline
0 & 1 & 1 & 0 & 0 & 1 & 1& 1& 2& \quad $PC=6$\\ \hline
2 & 2 & 2 & 0 & 1 &  $n$ & $n$& 2 & 2& \quad $PC=4$\\ \hline
2 & 2 & 0 & 1 & 1 & 0 & 2 & 2& 2& \quad $PC=3$\\ \hline \end{tabular}

{\bf Table 1:} The workings of the noncover $n$-algorithm

By $PC=2$ we mean that at this stage the {\it pending conjunction} is the second one, i.e. the one that defines $\mbox{Mod}_2$. In other words, we need to sieve out those $x \in \mbox{Mod}_1$ that happen to be in $\mbox{Mod}_2=(2,2,2,n,n,n,2,2,2)$. In order to do so we determine the intersection $\{-2,-1,0\} \cap \{-1, 0,1\} = \{-1,0\}$ of the ``$n$-pools'' of $\mbox{Mod}_1$ and $\mbox{Mod}_2$ and then split the $\{0,1,2,n\}$-{\it valued} row $r := \mbox{Mod}_1$ accordingly into a disjoint union $r = r' \cup r''$ where
$$\begin{array}{lllllll} r' & := & \{x \in r | \ x_{-1} =0 \ \mbox{or} \ x_0 =0 \}& =& (2,2,2,  {\bf n}, {\bf n}, 2,2,2,2)\\
r'' & := & \{x \in r | \ x_{-1} = x_0 =1 \} & =& (2,2,0,  {\bf 1}, {\bf 1}, 2,2,2,2). \end{array}$$
While all $x \in r'$ trivially satisfy $x_{-1} \wedge x_0 \wedge x_1 =0$, i.e. belong to $\mbox{Mod}_2$, this is not the case for all $x \in r''$. However, turning at the $6$-th position the 2 to 0 does the job. This yields the current {\it working stack} with the two rows labelled $PC= 3$; see top of Table 1. (Of course this ``stack'' has nothing to do with its namesake in ``stack filter''.)  As a general rule, the topmost row in the stack is always treated first (``last in, first out''). This may entail ``local changes'', or a splitting of the top row into several sons. In this way we proceed up to the second last stack in Table 1. Let us pick its top row $r = (2,n,n, 2, 0, n,n,n, 2)$ and illustrate once more the splitting process. The intersection of the $n$-pool of $r$ with (the index set of) the pending $6$th conjunction is $\{-3, -2, 1, 2, 3\} \cap \{-3, -2, -1, 2, 3, 4\} = \{-3, -2, 2, 3\}$. Accordingly split $r$ into the disjoint union of $r'$ and $r''$:
$$\begin{array}{lll}
r & = & (2, n, n, 2, 0, n, n, n, 2)\\
r' & = & (2, {\bf n}, {\bf n}, 2, 0, 2, {\bf n}, {\bf n}, 2)\\
r'' & = & (2, {\bf 1}, {\bf 1}, 2, 0, 0, {\bf 1}, {\bf 1}, 2). \end{array}$$
Since $r' \subseteq \mbox{Mod}_6$, $r'$ is the first son of $r$. We have $r'' \not\subseteq \mbox{Mod}_6$, but $r'' \cap \mbox{Mod}_6 = (2, 1, 1, n, 0, 0, 1, 1, n)$ becomes the second son. Both rows are {\it final}, i.e. are subsets of $\mbox{Mod}$ and thus collected in a steadily increasing {\it final stack}. The working stack now contains three rows with pending conjunctions $6, 4, 3$ respectively. In our case it just so happens that they are in fact already final (so e.g. {\it all} $x$ in the row labelled $PC = 4$ happen to satisfy the 4th, 5th and 6th conjunction). The final stack comprises thus the five rows in Table 2 (for the moment ignore $p^2q^2$ and so forth):

\hspace*{2cm} \begin{tabular}{|c|c|c|c|c|c|c|c|c|l}
$-4$ & $-3$ & $-2$ & $-1$ & 0 & 1 & 2 & 3 & 4& \\ \hline
2 & 2& 0 & 1 & 1& 0 & 2 & 2 &2& \quad $p^2q^2$\\ \hline
2 & 2 & 2& 0 & 1 & $n$ & $n$ & 2 & 2& \quad $pq(1-q^2) = pq-pq^3$ \\ \hline
0 & 1&1 & 0 & 0 & 1 & 1 & 1& 2 & \quad $p^3q^5$\\ \hline
2 & 1 & 1 & $n$ & 0 & 0 & 1 & 1 & $n$& \quad $p^2q^4(1-q^2)=p^2q^4-p^2q^6$\\ \hline
2 & $n$ & $n$ & 2 &0 & 2 & $n$ & $n$ &2 & \quad $p(1-q^4)=p-pq^4$\\ \hline \end{tabular}

{\bf Table 2:} The probability contributions of the final rows

For instance, the second row in Table 2 contains $2^5 \cdot (2^2-1)$ noncovers, where $(2^2-1)$ comes from $nn$. The total number $N$ of noncovers evaluates to

\hspace*{4.5cm} $N\quad = \quad 32+32 \cdot 3 + 2 + 2\cdot 3 + 16 \cdot 15 \quad = \quad 376,$ 

which is much higher than the number $R = 5$ of final multivalued rows. As we shall see in Section 3, in general the $n$-pool of rows is a bit more subtle.

Let us now calculate the output distribution.  The first row in Table 2 contains $2^5 = 32$ bitstrings $x$ with $b_1(x) =0$. Each contributes some probability $\alpha_1 \  \alpha_2 \ p \ q \ q \ p \ \alpha_3 \ \alpha_4 \ \alpha_5$ to the sum in (2). Since each $\alpha_i$ can independently be chosen to be $p$ or $q$,  the sum of these $32$ terms is
\begin{equation}
p^2q^2 (ppppp + \cdots + pqqpq+ \cdots + qqqqq)\quad =\quad p^2q^2 (p+q)^5 \quad =\quad p^2q^2.
\end{equation}
The fact that e.g. $nn = \{00, 01, 10\}$ yields  $pp +pq+qp = 1-q^2$, explains the contribution $pq(1-q^2)$ of the second row. Similarly for the three other rows.
Summing up the terms in Table 2 yields 
\begin{eqnarray}
\phi_S(p) & = & p^2q^2 +pq - pq^3+p^3q^5+p^2q^4-p^2q^6+p-pq^4 \nonumber \\
\nonumber \\ 
& = & 7p^2 - 8p^3-8p^4 + 25p^5-24p^6+11p^7 - 2p^8.
\end{eqnarray}

\section{Theoretic assessment}

Suppose the constraint $A^\ast = \{3,4\}$ is to be imposed on a row $r = (1,2,1,1)$ in the process of the stack filter $n$-algorithm. Then $r$ needs to be cancelled since {\it no} member $X \in r$ satisfies $X \not\supseteq A^\ast$. Fortunately, with some precautions the cancellation of rows can be avoided, which is essential in the Theorem below. Another remark about the proof is in order. Apart from the probabilities coupled to the final $\{0,1,2,n\}$-valued rows, the stack filter $n$-algorithm coincides with the noncover $n$-algorithm of [13], which is a special ``homogeneous'' case of the Horn $n$-algorithm, which in turn is an instance of some {\it principle of exclusion}. Since our special case is somewhat buried by this and the technical machinery of [13], yet admits a comparatively smooth proof from scratch, we give that proof below.

\begin{tabular}{|l|} \hline \\
{\bf Theorem:} Suppose the stack filter $S$ has window size $w$ and its positive Boolean function\\
 $b(x)$ is given as a disjunction of $h$ conjunctions (DNF). Then the stack filter $n$-algorithm \\
 computes the output distribution of $S$ in time $O(Nw^2h^2)$. Here $N$ is the number of \\
 bitstrings $x$ with $b(x)=0$. \\ \\ \hline \end{tabular}

{\it Proof.} As in the introductory example, the terms in the DNF of $b(x)$ yield subsets $A^\ast_1, \ldots, A^\ast_h$ of $W: = [w]$ whose models ($=$ simultaneous noncovers) $Y \subseteq W$ we wish to pack in disjoint $\{0,1,2,n\}$-valued rows. Any (not necessarily final) row $\ol{r}$ is called {\it feasible} if $Y \in \ol{r}$ for at least one model $Y$. As opposed to other applications of the princple of exclusion, here feasibility is easily tested. Namely, $\ol{r}$ is feasible if and only if
\begin{equation}
(\forall 1 \leq i \leq h) \quad A^\ast_i \not\subseteq \ \mbox{ones}(\ol{r}).
\end{equation}
Initially our ``working stack'' solely comprises the row $r_0 = (2,2,\ldots, 2)$ of length $w$ which we identify with the powerset of $W$. Note that $r_0$ is feasible since $\emptyset \in r$. Row $r_0$ carries the pointer $PC(r_0) =1$, where $PC$ stands for ``pending constraint''. Generally, the top row $r$ of the working stack is treated as follows. If $PC(r) = j$ (for some $j \in [h]$) then the set $A^\ast_j$ is ``imposed'' upon $r$, that is, the set $U$ of all $X \in r$ with $X \not\supseteq A^\ast_j$ is represented as a disjoint union of rows $r_1, \ldots, r_s$ where $s \leq w$. That  this is always possible (the ``core'' claim), and costs $O(w^2)$, will be shown in a moment. 

Because $r$ was feasible by induction, at least one of its ``candidate'' sons $r_1, \ldots, r_s$ will be as well. Since the feasibility of $\ol{r} = r_j$ amounts to the truth of (6), it costs $O(shw) = O(hw^2)$ to sieve the {\it sons} of $r$, i.e. the feasible rows among $r_1, \ldots, r_s$. Altogether the cost of one imposition of a constraint upon a row is $O(w^2)+ O(hw^2) =O(hw^2)$.

The $R$ final rows can be viewed as the leaves of a tree with root $(2,2,\ldots, 2)$ that has height $h$; each imposition triggers all sons of some node. Therefore the number of impositions is at most $Rh$ (distinct final rows, possibly having some of their forefathers in common). It follows that producing the $R$ final rows costs $O(Rh \cdot hw^2) = O(Nh^2w^2)$ in view of $R \leq N$, by the disjointness of final rows. Calculating (as in (4)) the contributions to $\phi_S(p)$ of all final rows, and adding them,  costs $O(Nw)$, which is swallowed by $O(Nh^2w^2)$.

It remains to verify the core claim, i.e. that $U: = \{X \in r: X \not\supseteq A^\ast \} \ (A^\ast:=A^\ast_j)$ can be represented as promised.

Case (a): \ $A^\ast\cap \ \mbox{zeros}(r) \neq \emptyset$ or $A^\ast$ wholly contains an $n$-bubble of $r$. Then $U  =r$, and so put $r_s = r_1 =r$.

Since $r$ is feasible, $A^\ast \subseteq \ \mbox{ones}(r)$ is impossible, and so the only remaining possibility is

Case (b): \ $A^\ast\cap \ \mbox{zeros}(r) = \emptyset$ and $A^\ast$ does not wholly contain an $n$-bubble of $r$ and $A^\ast \not\subseteq \ \mbox{ones}(r)$. This is exactly Case 7 in Section 5 of [13], whose essense we repeat here.

\begin{tabular}{|c|c|c|c|c|c|c|c|c|c|c|c|c|c|l}
$1$ & $2$ & $3$ & $4$ & $5$  & $6$ & $7$ & $8$ & $9$ & $10$ & $11$ & $12$ & $13$ & $14$ & \\ \hline 
$2$ & $2$ & $n_1$ & $n_1$ & $n_2$ & $n_3$ & $n_3$ & $n_4$ & $n_1$ & $n_2$ & $n_3$ & $n_3$ & $n_4$ & $n_4$ & \quad $r$ \\ \hline \hline
$2$ & $2$ & ${\bf n}$ & ${\bf n}$ & $n_2$ & $n_3$ & $n_3$ & $n_4$ & $2$ & $n_2$ & $n_3$ & $n_3$ & $n_4$ & $n_4$ & \quad $r_1$ \\ \hline
$2$ & $2$ & ${\bf 1}$ & ${\bf 1}$ & ${\bf 0}$ & $n_3$ & $n_3$ & $n_4$ & $0$ & $2$ & $n_3$ & $n_3$ & $n_4$ & $n_4$ & \quad $r_2$ \\ \hline
$2$ & $2$ & ${\bf 1}$ & ${\bf 1}$ & ${\bf 1}$ & ${\bf n}$ & ${\bf n}$ & $n_4$ & $0$ & $0$ & $2$ & $2$ & $n_4$ & $n_4$ & \quad $r_3$ \\ \hline
$2$ & $2$ & ${\bf 1}$ & ${\bf 1}$ & ${\bf 1}$ & ${\bf 1}$ & ${\bf 1}$ & ${\bf 0}$ & $0$ & $0$ & $n_3$ & $n_3$ & $2$ & $2$  & \quad $r_4$ \\ \hline
$n$ & $n$ & ${\bf 1}$ & ${\bf 1}$ & ${\bf 1}$  & ${\bf 1}$ & ${\bf 1}$ & ${\bf 1}$ & $0$ & $0$ & $n_3$ & $n_3$ & $n_4$ & $n_4$ & \quad $r_5$ \\ \hline \hline
\end{tabular}

{\bf Table 3:} Five candidate sons of some $\{0,1,2,n\}$-valued row

Suppose $W = [14],  \ A^\ast =[8]$, and $r$ is as in Table 3. Note that the $n$-pool of $r$ is the disjoint union $\{3,4,9\} \cup \{5,10\} \cup \{6,7,11, 12\}$ $\cup \{8,13, 14\}$ of four mutually independent $n$-{\it bubbles}, each one defined by ``at least one nul there'', as in Section 2. Putting

$r_1: = \{X \in r: \ X \not\supseteq \{3,4\}\}$

$r_2: = \{X \in r: \ X \supseteq \{3,4\} \ \mbox{and} \ X \not\supseteq \{5\} \}$

$r_3: = \{X \in r: \ X \supseteq \{3,4,5\} \ \mbox{and} \ X \not\supseteq \{6,7\} \}$

$r_4: = \{X \in r: \ X \supseteq \{3,4,5,6,7\} \ \mbox{and} \ X \not\supseteq \{8\} \}$

$r_5: = \{X \in r: \ X \supseteq \{3,4,5,6,7,8\} \ \mbox{and} \ X \not\supseteq \{1,2\} \}$

it is clear that $U$ is the disjoint union of $r_1, \ldots, r_5$. A minute's reflection shows that, crucially,  these sets can again be written as $\{0,1,2,n\}$-valued rows as shown in Table 3, and that generally (full details in [13]) splitting a row in $s \leq w$ candidate sons like this costs $O(sw) = O(w^2)$. \hfill $\square$

It follows from the proof that $O(Nw^2h^2)$ could be substituted by $O(Rw^2h^2)$ where $R \leq N$ is the number of final $\{0,1,2,n\}$-valued rows. Unfortunately $R$ is unpredictable. Theoretically $R=N$ is possible\footnote{Even then, provided $N \approx \frac{1}{2} \cdot 2^W$ as is to be expected for random PBF's, the stack filter algorithm would beat by a factor $2$ a brute force search of all of $\{0,1\}^W$.} but in practise $R$ is usually orders of magnitudes smaller than $N$ (see Subsection 4.1).

It has been pointed out that $b(x)$ may not initially be given in disjunctive normal form. However, if not, there are efficient methods to compute the DNF from any reasonable kind of presentation of $b(x)$; this e.g. applies to the erosion - dilation cascades in subsection 4.1. In any case, the bigger problem arguably is to find the bitstrings $x \in \{0,1\}^w$ with $b(x)=0$.

\section{Related matters}

Subsection 4.1 glimpses at the practical performance of a Mathematica implementation of the stack filter $n$-algorithm, and 4.2 shows that the so called rank selection probabilities $p_i$ can be gleaned from the $\{0,1,2,n\}$-valued rows along the way. Subsections 4.3 and 4.4 are about the joint distribution of stack filters, respectively about a certain generalization of ``ordinary'' stack filters to ``balanced'' stack filters. The required adaptions of our algorithm stay within the realm of $\{0,1,2,n\}$-valued rows. Finally, as powerful as binary decision diagrams often are, it is argued in 4.5 that they are not appropriate in our situation.

\subsection{Numerics exemplified on the $LULU$ filter $C_5$}

Certain stack filters $L_n$, their duals $U_n$, and compositions thereof (called $LULU$ filters) have been proposed in [9] and earlier\footnote{Using terminology of Mathematical Morphology, $L_n$ (dually for $U_n$) is an opening induced by a line segment of length $n+1$, whence the underlying PBF has window size $2n+1$. As opposed to the median filters, all $LULU$ filters $S$ are idempotent $(S\circ S=S)$ and even co-idempotent $((id-S) \circ (id-S) =id-S)$.}, as alternatives to the popular median filters. 
Actually,  the function $b_1 (x_{-4}, \ldots, x_4)$ from Section 2 is the PBF underlying $U_2L_2$.  

The natural definition of each $LULU$ filter is as a {\it cascade} of so called {\it erosions} and {\it dilations} (CED), two dual concepts from Mathematical Morphology [9, III.C].  Computing the DNF of any CED essentially amounts\footnote{Computing the DNF of a PBF from its CNF is a well researched topic [5], which also amounts to get all minimal transversals of a set system. The author used a refinement of  the classic ``Berge-algorithm'' for the task, does not claim that it competes with the cutting edge algorithms for DNF $\leftrightarrow$ CNF, but feels that the stack filter $n$-algorithm is the right approach once the DNF is {\it given}.} to calculating
CNF's and DNF's of successively bigger (details in [3]) positive Boolean functions. 
For instance, 
$$C_n := L_nU_nL_{n-1}U_{n-1} \ldots L_1U_1$$
is a CED stack filter with window size $w = 2n^2+2n+1$. Using Berge's algorithm to compute the DNF of $C_5$ from its CED-representation took about 46 hours. Calculating the output distribution $\phi_{C_5} (p) = p^5+7p^6- \cdots + 114680p^{43} + \cdots + p^{53}$ with the stack filter $n$-algorithm took another 12 hours. At least as illuminating as $\phi_{C_5} (p)$ are the so called rank selection probabilities that can be calculated along the way as discussed in the next subsection. The underlying PBF of $C_5$ had $N = 639'173'390'187'370'752$ models, which were packed in a mere $R = 179'244$ final rows. More extensive numerical evaluations of similar implementations of the principle of exclusion (and how they compare to say BDD's) are provided in upcoming publications.

Due to the specific regularities of $U_nL_n$ its DNF has in fact been discovered by other means  [9, p.112] and its distribution transfer was computed independent of its DNF in [3]; it equals

\hspace*{3cm} $\phi_{U_nL_n}\quad = \quad 1-q^{n+1} - npq^{n+1} - pq^{2n+2} - \frac{1}{2}(n-1)(n+2)p^2q^{2n+2}$. \hfill (7)

One verifies that (7) coincides with (5) for $n =2$. Even the distribution transfer of $C_n$ can be determined  [3], albeit only by an efficient recursive formula as opposed to the closed form in (7). For all $n \leq 5$ the results agreed with the ones obtained with the stack filter $n$-algorithm, which is a strong indication that both methods are correct.

\subsection{Rank selection probabilities}

Let $S$ be a stack filter.  Given a sequence $Z$ of i.i.d. random variables, the so called {\it rank selection probability} $p_i$ is defined as the probability that a fixed component of the output series $SZ$ is the $i$-th smallest in the sliding window of length $w$. It is  known, [8], [2, p.236] that
$$p_i \quad =\quad \frac{A_{w-i}}{{w \choose w-i}} - \frac{A_{w-i+1}}{{w \choose w-i+1}},$$
where $A_i$ is the number of bitstrings $x$ with $i$ ones and $w-i$ zeros that have $b(x) =0$. The $A_i$'s can be conveniently calculated in tandem with the evaluation of (2). For instance, as the reader can easily verify, the contribution of the last row in Table 2 to $A_0$ up to $A_7$ is:
$$\begin{array}{rrrllll}
A_0 & : & {8 \choose 0} & = & 1\\ \\
A_1 & : & {8 \choose 1} & = & 8 \\ \\
A_2 & : & { 8 \choose 2} &= & 28 \\ \\
A_3 & : & { 8 \choose 3} & = & 56 \\
\\
A_4 &: & {4 \choose 0}{4 \choose 4} + {4 \choose 1}{4 \choose 3} + {4 \choose 2}{4 \choose 2} + {4 \choose 3} { 4 \choose 1} & = & 69\\ \\
A_5 & : & {4 \choose 1} {4 \choose 4} + {4 \choose 2}{4 \choose 3} + {4 \choose 3}{4 \choose 2} &= & 52\\
\\
A_6 &: & {4 \choose 2}{4 \choose 4} + {4 \choose 3}{4 \choose 3} &= & 22\\
\\
A_7 &:& {4\choose 3}{4 \choose 4} &= & 4. \\
\end{array}$$

We mention that in [6] the optimization of stack filters with respect to certain constraints leads to specific desirable values of $A_1, \ldots, A_w$. Finding a stack filter $S$ that features these values (at least approximately) is however hard. One may hence be led to compile a catalogue of CED's (see 4.1) with corresponding vectors $(A_1, \ldots, A_w)$ from which a suitable candidate $S$ can be picked.

\subsection{The joint output distribution of two stack filters}

Let $Z$ be a doubly infinite sequence of i.i.d. random variables. For two stack filters $S$ and $T$ with corresponding positive Boolean functions $b_1(x)$ and $b_2(y)$ their joint output distribution $F_{SZ, TZ}(s,t)$, or simply $JD(s,t)$, is defined as
$$JD(s,t)\quad : =\quad \ \mbox{Prob}((SZ)_0 \leq s \ \mbox{and} \ (TZ)_0 \leq t).$$
If we set $p := \ \mbox{Prob}(Z_0 \leq s), \ \pi : = \ \mbox{Prob}(Z_0 \leq t)$ and assume $p \leq \pi$ (the case $p > \pi$ is similar) then it is shown in [2, p.230]  that

\hspace*{4cm} $JD(s,t)\quad =\quad \ds\sum_{i=0}^w \ds\sum_{j=0}^w A_{i,j} p^i(\pi -p)^{w-i-j}(1-\pi)^j$, \hfill (8)

where $A_{ij}$ is the number of  $(x,y) \in \{0,1\}^w \times \{0,1\}^w$     such that
$$x \geq y, \quad b_1(x) = b_2(y) =0, \quad v_{-,-}(x,y)=i, \quad v_{+,+}(x,y) = j,$$
and where\footnote{This notation is not used in [2] but ties in well with the notation used in subsection 4.4, which in turn is akin to the notation of [11]. For instance, our $v_{-,+}(x,y)$ in 4.4 corresponds to $w(\ol{x}\wedge s)$ in equation (17) of [11].}
$$\begin{array}{lll}
v_{-,-} (x_1, \ldots, x_w, y_1, \ldots, y_w) &  : = & |\{1 \leq k \leq w: x_k = y_k = 0 \}| \\
\\
v_{+,+} (x_1, \ldots, x_w, y_1, \ldots, y_w) &:= & |\{ 1 \leq k \leq w: x_k = y_k =1\}| \end{array}$$
The calculation of the coefficients $A_{ij}$ works row-wise. So suppose $r$ in Table 4 is one of the final rows obtained after applying the noncover $n$-algorithm to $b_1$.  Obviously the set
$${\cal F}\quad := \quad \{y: (\exists x \in r) \ \ x \geq y\}$$
is represented by row $r_0$. If say $b_2(y) = y_3 \wedge y_9 \wedge y_{10}$ then the set
$${\cal F}(b_2) \quad : = \quad \{y \in {\cal F}: b_2(y) = 0\}$$
is the disjoint union $\rho_1 \cup \rho_2 \cup \rho_3$:

\hspace*{1cm} \begin{tabular}{l|c|c|c|c|c|c|c|c|c|c|c|}
 & 1 & 2 & 3& 4 &5 &6 &7 & 8 &9 & 10 & 11 \\ \hline
$r=$ & $n_1$ & $n_1$ & $n_1$ & 2 & 2 & 0 & $n_2$ & $n_2$ & $n_2$ & 1 &1 \\ \hline
$r_0=$ & $n_1$ & $n_1$ & $n_1$ & 2 & 2 & 0 & $n_2$ & $n_2$ & $n_2$ & 2 &2 \\ \hline
$\rho_1=$ &2 &2 & ${\bf 0}$ &2 & 2 & 0 & $n_2$ & $n_2$ & ${\bf n_2}$ & ${\bf 2}$ & $2$ \\ \hline 
$\rho_2 =$ & $n_1$ & $n_1$ & ${\bf 1}$ & 2 & 2 & 0 & 2 & 2 & ${\bf 0}$ & ${\bf 2}$ & $2$ \\ \hline 
$\rho_3=$ & $n_1$ & $n_1$ & ${\bf 1}$ & 2 & 2& 0 & $n_2$ & $n_2$ & ${\bf 1}$ & ${\bf 0}$ &2 \\ \hline 
$x=$ & 0 & 1 & 1 & 1&0 & 0 & 1 & 1 &0 & 1 & 1 \\ \hline
$\sigma =$ & 0 & 2 &2 &2 &0 & 0 & ${\bf 0}$ & ${\bf 2}$ & 0 & 0 &2 \\ \hline
$\tau =$ & 0 & 2 & 2 & 2 & 0 & 0 & ${\bf 1}$ & ${\bf 0}$ &0 & 0 & 2\\ \hline
\end{tabular}

{\bf Table 4:} Adapting the algorithm to joint output distributions

For each $x \in r$ and $k \in \{1,2,3\}$ one now records $v_{-,-}(x,y)$ and $v_{+,+}(x,y)$ for all $y \in \rho_k$ with $y \leq x$. For instance, taking the $x$ indicated in Table 4 one verifies that
$$\{y \in \rho_3 : \ y \leq x\}\quad = \quad\sigma \cup \tau,$$
where the later union is disjoint (see $n_2n_2$ in $\rho_3$ and the corresponding boldface entries in $\sigma, \tau$). It is easy to see that $\sigma$ contributes an amount of ${5 \choose j}$ to the value of $A_{4,j}$ for all $0 \leq j \leq 5$. Similarly $\tau$ contributes an amount of ${4 \choose j}$ to the value of $A_{4, j+1} \ (0 \leq j \leq 4)$. Calculations can be sped up by clumping together suitable $x$'s rather than processing them one by one. We discuss a similar phenomenon in more detail in the next subsection.

\subsection{Balanced stack filters}

In [11], [12] the concept of a {\it balanced}\footnote{Actually, Arce, Paredes and Shmulevich propose to reserve the term ``stack filter'' to their new concept, and to relabel the ``old'' stack filters as stack smoothers. As suggested by one referee, we stick to the old, well established terminology.} stack filter $S$ is introduced. Citing from [11]: ``They are much more versatile, being empowered not only with lowpass filtering characteristics, but with bandpass or highpass filtering characteristics as well.'' They are based on ``mirrored thresholding'' which entails $t$ and $-t$ to play symmetric roles. Most important for us, $S$ is based again upon a PBF albeit in a manner more sophisticated than (1). For instance, the PBF is of the kind $b(x,y) = b(x_1, \ldots, x_w, y_1, \ldots, y_w)$, and in this set up a stack filter turns out to be a balanced stack filter where $b$ does not depend on $y_1, \ldots, y_w$ (i.e., these variables are fictitious). As usual let $Z$ be a doubly inifinite sequence of i.i.d. random variables with common cumulative distribution function $F_Z(t) = Prob (Z_i \leq t) \ (i \in \Z)$. Put $F(t) = F_Z(t)$ and

$\begin{array}{lll}
p_{+,+} &  : = & \left\{ \begin{array}{lll} F(-t) -F(t) & \mbox{if} & t \leq 0 \\ 0 & \mbox{if} & t > 0 \end{array}\right.\\
\\
p_{-,-} & := & \left\{ \begin{array}{lll} 0 & \mbox{if} & t \leq 0 \\
F(t) - F(-t) & \mbox{if} & t > 0 \end{array}\right. \\
\\
 p_{-, +} & : = & \left\{ \begin{array}{lll} F(t) & \mbox{if} & t \leq 0 \\
F(-t) & \mbox{if} & t > 0 \end{array} \right.\\
\\
p_{+, -} & : = & \left\{ \begin{array}{lll} 1 - F(-t) & \mbox{if} &t  \leq 0 \\
1 - F(t) & \mbox{if} & t > 0. \end{array} \right. \end{array}$

Besides $v_{+,+}(x,y)$ and $v_{-,-}(x,y)$ from 4.3 we also put
$$\begin{array}{lll}
v_{-,+}(x,y) &: = & |\{1 \leq k \leq w: \ x_k = 0 \ \mbox{and} \ y_k = 1 \} | \\
\\
v_{+, -}(x,y) & : = & | \{1 \leq k \leq w: \ x_k =1 \ \mbox{and} \ y_k =0 \} |. \end{array}$$
Modulo some obvious typos, it  is shown in [11, (17)] that the output distribution, i.e. $F_{SZ}(t) = Prob((SZ)_0 \leq t)$, can be calculated as
$$F_{SZ}(t)\quad =\quad \ds\sum_{b(x,y)=0} p^{v_{+,+}(x,y)}_{+,+} \cdot p^{v_{+,-}(x,y)}_{+,-} \cdot p^{v_{-,+}(x,y)}_{-,+} \cdot p^{v_{-,-}(x,y)}_{-,-}.$$
As opposed to $JD(s,t)$ in (8), which is a polynomial of Prob$(Z_0\leq s)$ and Prob$(Z_0 \leq t)$, here $F_{SZ}(t)$ is not quite a polynomial in terms of $Prob((SZ)_0 \leq t)$ and $Prob((SZ)_0 \leq -t)$. 

Nevertheless the noncover $n$-algorithm is of good use. Suppose it has (among others) returned the final row $r$ in Table 5. 
Take any bitstring $x^\ast = (x_1, \ldots, x_9)$ ``contained'' in the left hand side $(n_1, n_2, n_3, 1, n_4, n_4, 0,2, n_3)$ of $r$. More precisely, any bitstring $x^\ast$ which is {\it extendible}\footnote{It is easily seen that the extendible bitstrings are exactly the members of $(2,2,2,1,n_4, n_4, 0,2,2)$.} to a bitstring $(X^\ast, y) \in r$.
Say $x^\ast = (1,1,1,1,1,0,0,0,0)$. For each fixed $k \in \{0,1, \ldots, 5\}$ and $k' \in \{0,1, \cdots, 4\}$ we now show how the number $f(k,k')$ of bitstrings $y = (y_1, \ldots, y_9)$ with
$$v_{+,+}(x^\ast,y) =k \quad \mbox{and} \quad v_{-,+}(x^\ast,y)  = k'$$
$$(\mbox{whence} \ v_{+, -}(x^\ast, y) = 5-k \quad \mbox{and} \quad v_{-,-}(x^\ast,y) = 4-k')$$
can be calculated fast. First, notice that the subset
$$r(x^\ast) \quad : =\quad \{(x,y) \in r: \ x = x^\ast\}$$
of $r$ can be written as multi-valued row as shown in Table 5.

\begin{tabular}{l|c|c|c|c|c|c|c|c|c|c|c|c|c|c|c|c|c|c|c|} 
& $x_1$ & $x_2$ & $x_3$ & $x_4$ & $x_5$ & $x_6$ & $x_7$ & $x_8$ & $x_9$ & \quad & $y_1$ &$y_2$ & $y_3$ & $y_4$ & $y_5$ & $y_6$ & $y_7$ & $y_8$ & $y_9$ \\ \hline 
& & & & & & & & & & & & & & & & & & & \\ \hline
$r=$ & $n_1$ & $n_2$ & $n_3$ & $1$ & $n_4$ & $n_4$ & $0$ & $2$ & $n_3$ & \quad & $n_1$ & $n_1$ & $n_3$ & $n_2$ & $n_2$ & $n_1$ & $n_1$ & $n_2$ & $n_2$ \\ \hline
$r(x^\ast) =$ & 1 & 1 & 1 & 1 &1  & 0 & 0 & 0 & 0 & \quad & $n_1$ & $n_1$ & $2$ & $n_2$ & $n_2$ & $n_1$ & $n_1$ & $n_2$ & $n_2$ \\ \hline
& & & & & & & & & & & & & & & & & & & \\ \hline
$r_1=$ & 1 & 1 & 1 & 1 & 1 &0 &0 & 0 &0 & \quad & ${\bf n_1}$ & ${\bf n_1}$ & $2$ & ${\bf n_2}$ & ${\bf n_2}$ & 2 & 2 & 2 & 2\\ \hline
$r_2=$ & 1 &1 &1 &1 &1 &0 &0 &  0 & 0 & \quad & ${\bf n_1}$ & ${\bf n_1}$ & 2 & {\bf 1} & {\bf 1} &2 & 2 & $n_2$ & $n_2$ \\ \hline
$r_3=$ & 1 & 1 & 1& 1& 1 &0 & 0 & 0 & 0 & \quad & {\bf 1} & {\bf 1} & 2 & ${\bf n_2}$ & ${\bf n_2}$ & $n_1$ & $n_1$ & 2 &2 \\ \hline
$r_4=$ & 1 &1 &1 &1 &1 & 0 & 0 & 0 & 0 & \quad & {\bf 1} & {\bf 1} & 2 & {\bf 1} & {\bf 1} & $n_1$ & $n_1$ & $n_2$ & $n_2$ \\ \hline
\end{tabular}

{\bf Table 5:} Adapting the algorithm to balanced stack filters

Problem is we  cannot freely choose $k$ 1's among $\{y_1, \ldots, y_5\}$ and $k'$ 1's among $\{y_6, \ldots, y_9\}$ because e.g. the choice $(1,1,0,0,0,1,1,0,0)$ clashes with $n_1n_1n_1n_1$. But when one partitions $r(x^\ast)$ as $r_1 \cup r_2 \cup r_3 \cup r_4$ as indicated, then for each $r_i$ the choices within $\{y_1, \ldots, y_5\}$ respectively $\{y_6, \ldots, y_9\}$ can be made independently. To fix ideas, say $k=2$ and $k'=3$. Then the contribution of $r(x^\ast) = r_1 \cup r_2 \cup r_3 \cup r_4$ to the coefficient of the monom
$$p_{+,+}^k \ p_{+,-}^{5-k} \ p^{k'}_{-,+} \ p^{4-k'}_{-,-}$$
occuring in $F_{SX}(t)$ is 
$$f(k,k')\quad =\quad 8 \cdot 4 \quad + \quad 1 \cdot 2 \quad  + \quad 1 \cdot 2 \quad + \quad 0 \cdot 0\quad = \quad 36.$$

Generally, the number of bitstrings with a fixed number $k$ of $1$'s that are contained in a $\{0,1,2,n\}$-valued row can be determined fast. Similar to 4.3, but more obvious, time can be saved by clumping together suitable bitstrings $(x_1, \ldots, x_9)$. For instance, $(1,1,0,1,0,0,0,1,1)$ causes the same right hand side $(n_1, n_1, 2, n_2, n_2, n_1, n_1, n_2, n_2)$ as did $x^\ast$. As another example, $(0,0,1,1,1,0,0,0,0)$ is one among ten left hand sides of weight 3 that cause the right hand side $(2,2,2,2,2,2,2,2,2)$.

\subsection{On binary decision diagrams}

Shmulevich et al. [10] proposed to evaluate (2) by setting up a binary decision diagram (BDD) for the Boolean function $b(x)$ that underlies the stack filter $S$ whose distribution transfer needs to be calculated.  Suppose one has indeed spent time to get a BDD that represents $b(x)$.  While the {\it number} of models $x \in \{0,1\}^w$ with $b(x) =0$ can be determined fast from a BDD, it is more cumbersome to {\it generate} all models, as is forced by (2). True, from the BDD one can get the set of models as a disjoint union of $\{0,1,2\}$-valued rows in recursive fashion. (See [1, p.22] or the long chapter on BDDs in Donald Knuth's forthcoming book.) However, these rows are far more numerous than the ones produced by the stack filter $n$-algorithm; not surprisingly since our algorithm uses one {\it additional} symbol and hence more flexibililty in its $\{0,1,2,n\}$-valued rows.
Finally, the enhancements discussed in subsections 4.2, 4.3, 4.4 are cumbersome to be handled by BDD's.

\section*{Conclusion}

The present article can be viewed as the realization of a fifth benefit of DNF's that was announced in [14], i.e. the calculation of a stack filter's output distribution and (even more useful) its selection probabilities. The so doing stack filter $n$-algorithm is accessible from the author's home page. It has the form of a Mathematica Notebook. The indicated enhancements in 4.3 and 4.4 have not been programmed by the author; anybody is welcome to do so.

Last not least we draw attention to [7], a comprehensive framework in which stack filters, alias {\it lattice polynomial functions} (LPF), constitute but one type of aggregation function. However, there are no references to nonlinear signal theory or Mathematical Morphology in [7]. For instance, other than might appear from [7, p.361], cumulative distribution functions of ``nice'' LPF's (i.e. their underlying PBF's are more regular than ours) have a long history - in the case of Order Statistics dating back to 1932 [4].

\end{document}